\documentclass[twocolumn,aps,prl,showpacs,superscriptaddress,groupedaddress]{revtex4}
\usepackage[latin9]{inputenc}
\usepackage{amsbsy}
\usepackage{amstext}
\usepackage{amssymb}
\usepackage{graphicx}
\usepackage{esint}

\makeatletter
\@ifundefined{textcolor}{}
{%
 \definecolor{BLACK}{gray}{0}
 \definecolor{WHITE}{gray}{1}
 \definecolor{RED}{rgb}{1,0,0}
 \definecolor{GREEN}{rgb}{0,1,0}
 \definecolor{BLUE}{rgb}{0,0,1}
 \definecolor{CYAN}{cmyk}{1,0,0,0}
 \definecolor{MAGENTA}{cmyk}{0,1,0,0}
 \definecolor{YELLOW}{cmyk}{0,0,1,0}
}

\usepackage{dcolumn}% needed for some tables
\usepackage{bm}% for math
\makeatother

\begin{document}
% The following information is for internal review, please remove them for submission%\widetext%\leftline{Version xx as of \today}%\leftline{Primary authors: Joe E. Physics}%\leftline{To be submitted to (PRL, PRD-RC, PRD, PLB; choose one.)}%\leftline{Comment to {\tt d0-run2eb-nnn@fnal.gov} by xxx, yyy}%\centerline{\em D\O\ INTERNAL DOCUMENT -- NOT FOR PUBLIC DISTRIBUTION}

% the following line is for submission, including submission to the arXiv!!%\hspace{5.2in} \mbox{Fermilab-Pub-04/xxx-E}

\title{Thermally driven topology in chiral magnets}

\author{Wen-Tao Hou}

\affiliation{Department of Physics, Boston College, Chestnut Hill, Massachusetts
02467, USA}
\affiliation{Department of Physics, University of New Hampshire, Durham, New Hampshire
03824, USA}

\author{Jie-Xiang Yu}

\affiliation{Department of Physics, University of New Hampshire, Durham, New Hampshire
03824, USA}

\author{Morgan Daly}

\affiliation{Department of Physics, University of New Hampshire, Durham, New Hampshire
03824, USA}

\author{Jiadong Zang}
\email{Jiadong.Zang@unh.edu}

\affiliation{Department of Physics, University of New Hampshire, Durham, New Hampshire
03824, USA}

\date{\today}
\begin{abstract}
Chiral magnets give rise to the anti-symmetric Dzyaloshinskii-Moriya
(DM) interaction, which induces topological nontrivial textures such
as magnetic skyrmions. The topology is characterized by integer values
of the topological charge. In this work, we performed the Monte-Carlo
calculation of a two-dimensional model of the chiral magnet. A surprising
upturn of the topological charge is identified at high fields and
high temperatures. This upturn is closely related to thermal fluctuations
at the atomic scale, and is explained by a simple physical picture
based on triangulation of the lattice. This emergent topology is also
explained by a field-theoretic analysis using $CP^{1}$ formalism. 
\end{abstract}
\maketitle
%\section{\label{sec:level1}First-level heading}% sections are not used for PRL papers

The marriage of topology and condensed matter physics has given birth
to numerous excitements in the past decades. In particular, magnetism,
the zoo of topological spin textures, such as domain walls, vortices
and Bloch points, not only gives rise to rich physics, but also leads
to transformative spintronics applications. The recently discovered
magnetic skyrmion is a new member of such topological textures\cite{Bogdanov1994,Muhlbauer_2009,Yu_2010_nat,Yu_2010_nmat}.
It is a two dimensional (2D) whirlpool-like structure with spins therein
pointing to all directions. It has one-to-one correspondence to the
three dimensional monopole defect by the stereographic mapping. Topology
of the skyrmion can be captured by the topological charge (TC)\cite{Rajaraman,Nagaosa2013}
\begin{equation}
Q=\frac{1}{4\pi}\int d^{2}r{\bf n}\cdot(\partial_{x}{\bf n}\times\partial_{y}{\bf n})\label{eq:TQ},
\end{equation}
where ${\bf n}$ is a unit vector describing the local spin direction.
It is valued $\pm1$ for each skyrmion, and cannot be altered by slight
deformation of the texture configuration. As a result of this nontrivial
topology, the skyrmion acquires novel properties, such as the topological
Hall effect and the skyrmion Hall effect\cite{Neubauer_2009,Yi_2009,Zang_2011,Litzius2017,Jiang2017},
which have potential in future topological devices\cite{Fert2013}. 

The magnetic skyrmion was originally proposed theoretically in noncentrosymmetric
magnets\cite{Bogdanov1994,Bogdanov1989a,Bogdanov1989b,Rossler} and
its crystal form was recently discovered in bulk sample of MnSi, a
typical family of noncentrosymmetric magnets, by small angle neutron
scattering\cite{Muhlbauer_2009}. It was later confirmed in (FeCo)Si
thin film by real space imaging with Lorentz transmission electron
microscopy\cite{Yu_2010_nat}. The skyrmion crystal phase in the thin
film is greatly extended in the $B$-$T$ diagram (where $B$ is magnetic
field and $T$ is temperature) compared to the bulk sample, which
has been further addressed by follow-up experiments\cite{Yu_2010_nmat}. 
This is because of the suppression of the conical phase in thin films.
But nevertheless, skyrmions still exist only below the Curie temperature.

In the skyrmion crystal phase, the TC is significant and essentially
counts the number of skyrmions therein. But TC in Eq.{[}\ref{eq:TQ}{]}
respects the rotational symmetry, so that it cannot serve as an order
parameter, and does not have correspondence to the crystal phase.
It is interesting to study the distribution of TC in the same $B$-$T$
phase diagram. To this end, we used the Monte Carlo method in this
work and studied the distribution of the TC. It is significantly
extended compared to the skyrmion crystal phase, and can be explained
by $CP^{1}$ modeling\cite{CP1-skyrmion,Auerbach}.

We studied a 2D film of chiral magnet, whose Hamiltonian is described
by the following classical spin model 
\begin{equation}
H=\sum_{\langle i,j\rangle}(-J{\bf S}_{i}\cdot{\bf S}_{j}+{\bf D}_{ij}\cdot{\bf S}_{i}\times{\bf S}_{j})-g\mu_{B}H'\sum_{i}S_{i}^{z},
\end{equation}
where ${\bf S}_{i}=S{\bf n}_{i}$ is the spin on site $i$ with ${\bf n}_{i}$,
a three dimensional unit vector, and $\langle i,j\rangle$ means the
nearest neighbors. In the Monte Carlo calculation, $S=1$ and a square
lattice is employed. $J>0$ is the ferromagnetic Heisenberg exchange
coupling, while ${\bf D}_{ij}$ is the vector of the DM interaction
between neighboring sites $i$ and $j$. The strength of DM interaction is $D=|{\bf D}_{ij}|$. The last term describes the
Zeeman coupling, where $\mu_{B}$ is the magnetic moment and $H'$
is the applied magnetic field along $z$ direction. We define $B=g\mu_{B}H'$
and choose the natural units ($\hbar=k_{B}=c=1$). It has been confirmed
that this simple Hamiltonian captures most essential physics of 2D
chiral magnets\cite{Yu_2010_nat,CP1-skyrmion,Dupe_2016}.

\begin{figure}
\includegraphics[scale=0.4]{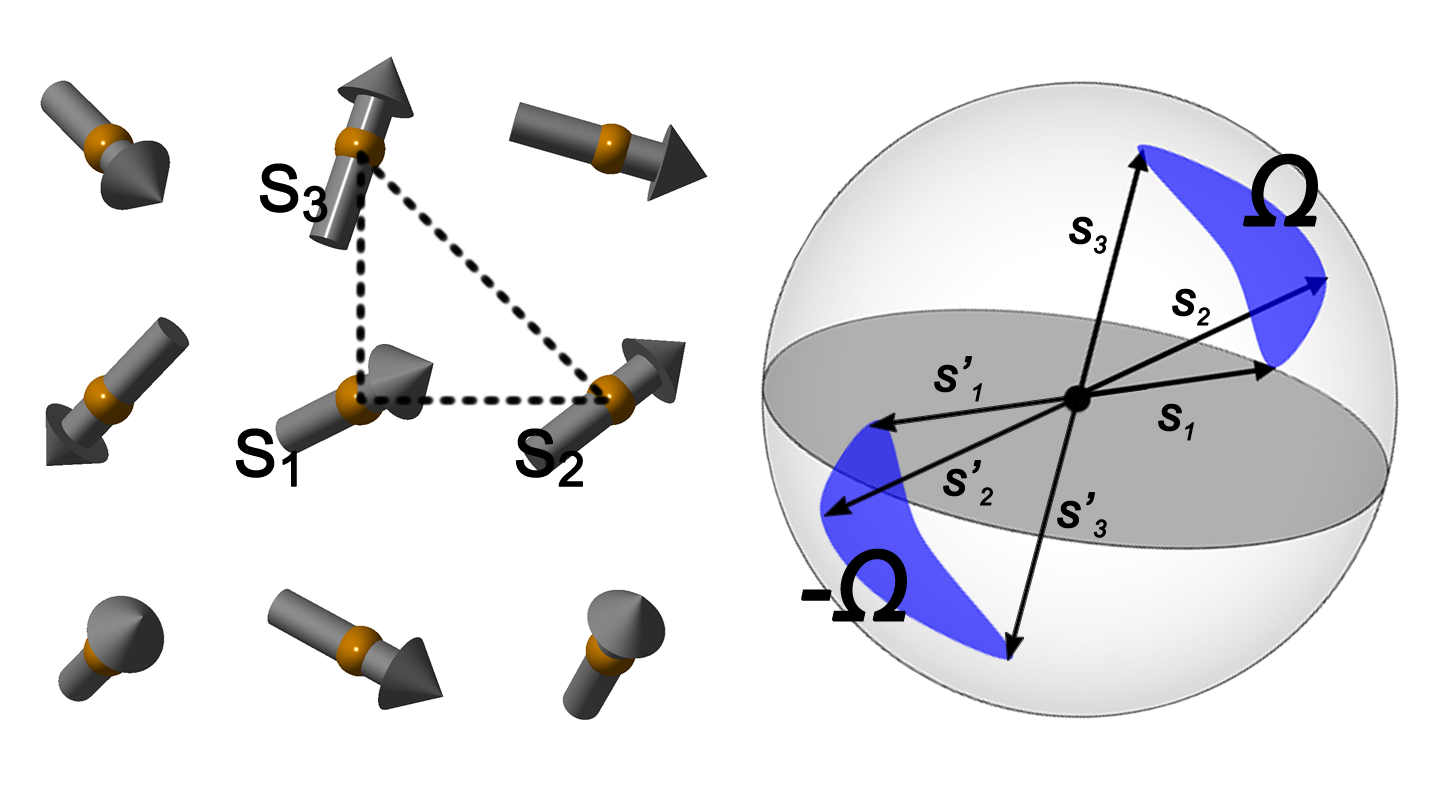} \caption{Schematic diagram of the TC obtained by the solid angle $\Omega$
for each three nearby spins ${\bf S}_{1}$,${\bf S}_{2}$ and ${\bf S}_{3}$.
This solid angle flips sign when the three spins are reversed to ${\bf S}_{1}'$,
${\bf S}_{2}'$ and ${\bf S}_{3}'$.}
\label{fig:lattice} 
\end{figure}

\begin{figure}
\includegraphics[scale=0.50]{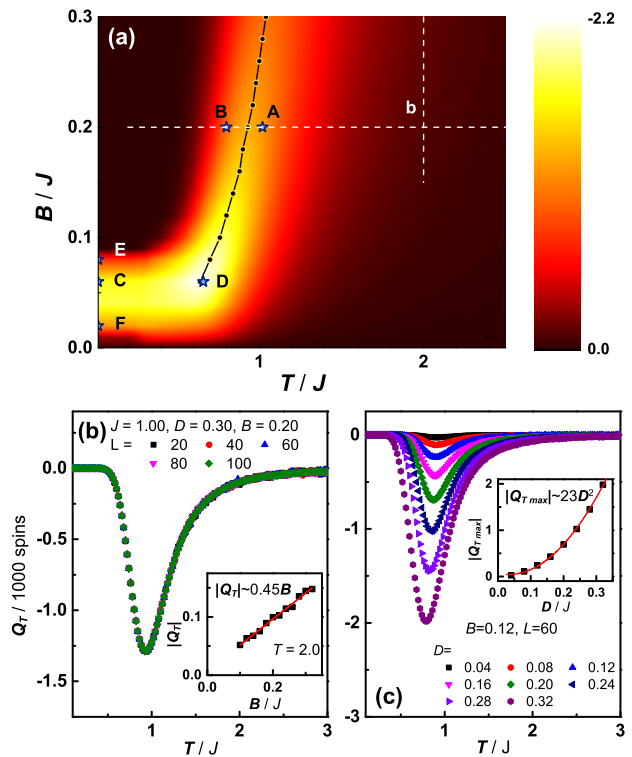}\caption{Field-, temperature- and DM- dependent TC. (a) The phase diagram of
TCs with the magnetic field and temperature dependence with $D=0.30J$
. The peak value is connected as a symbolled line. The horizon and
vertical dashed lines correspond to the finite size test in (b) and
field-dependent TC in the insert panel of (b). Star symbols labeled
A to F correspond the snapshot in Fig. 3. (c) The TC as a function
of DM interaction with the fixed magnetic field $B=0.12J$. The insert
panel shows the square relationship between the peak value of TC and
DM interaction. }
\label{fig:phase} 
\end{figure}

To calculate the thermal average of the TC, we triangulated the square
lattice. Summation over all the solid angles $\Omega$ of three spins
on each triangle divided by $4\pi$ gives the total TC for each spin
configuration. $\Omega$ is computed by the Berg formula\cite{Berg_1981}:
\begin{equation}
\exp(\frac{i\Omega}{2})=\rho^{-1}[1+{\bf n}_{1}\cdot{\bf n}_{2}+{\bf n}_{2}\cdot{\bf n}_{3}+{\bf n}_{3}\cdot{\bf n}_{1}+i{\bf n}_{1}\cdot({\bf n}_{2}\times{\bf n}_{3})],\label{eq:berg}
\end{equation}
where ${\bf n}_{1}$, ${\bf n}_{2}$ and ${\bf n}_{3}$ are three
spins on the triangle and $\rho=[2(1+{\bf n}_{1}\cdot{\bf n}_{2})(1+{\bf n}_{2}\cdot{\bf n}_{3})(1+{\bf n}_{3}\cdot{\bf n}_{1})]^{\frac{1}{2}}$
is the normalization factor. The Metropolis \cite{Metropolis_1949}
and over-relaxation algorithm are employed iteratively to generate
a Markov chain of spin configurations\cite{Metropolis_1949,Creutz1987},
averaging over which thermal average of the TC was derived. We imposed
periodic boundary conditions and performed averaged over $2.4\times10^{6}$
ensembles at each temperature. The main results of the TCs are shown
in Fig. \ref{fig:phase}(a). It shows the color plot of the average
TC in the $B$-$T$ diagram with the fixed DM interaction as $D=0.3J$.
A dramatic upturn of the TC is addressed along a ridge in the phase
diagram. The value of the TC is significant in areas greatly extended
to the skyrmion phase, which is located at small $B$ and low $T$
in the bottom region of the ridge.

Special attentions are paid to the high field region, where no skyrmions
are expected. As a typical example, we fix the field at $B=0.2J$,
and the relation between average TC and temperature is shown in Fig.
\ref{fig:phase}(b). At very low temperature, TC is equal to zero,
as all spins are nearly polarized. At very high temperature, TC again
converges to zero due to the topological triviality of a completely
random phase. However, in between, TC becomes significantly elevated
at finite temperatures. A deep dip of the TC is witnessed around $T=1.0J$,
the Curie temperature of the corresponding Heisenberg model. Here,
the negative TC is consistent with the fact that the spin at the skyrmion
core is opposite to the external magnetic field. The same calculations
were performed for lattices with sizes varying from $20\times20$
to $100\times100$. No difference could be found between different
lattice sizes. This immunity to the finite size effect suggests robustness
of the TC upturn, which might be related to the scaling-free atomic
scale physics.

\begin{figure}
\includegraphics[scale=0.16]{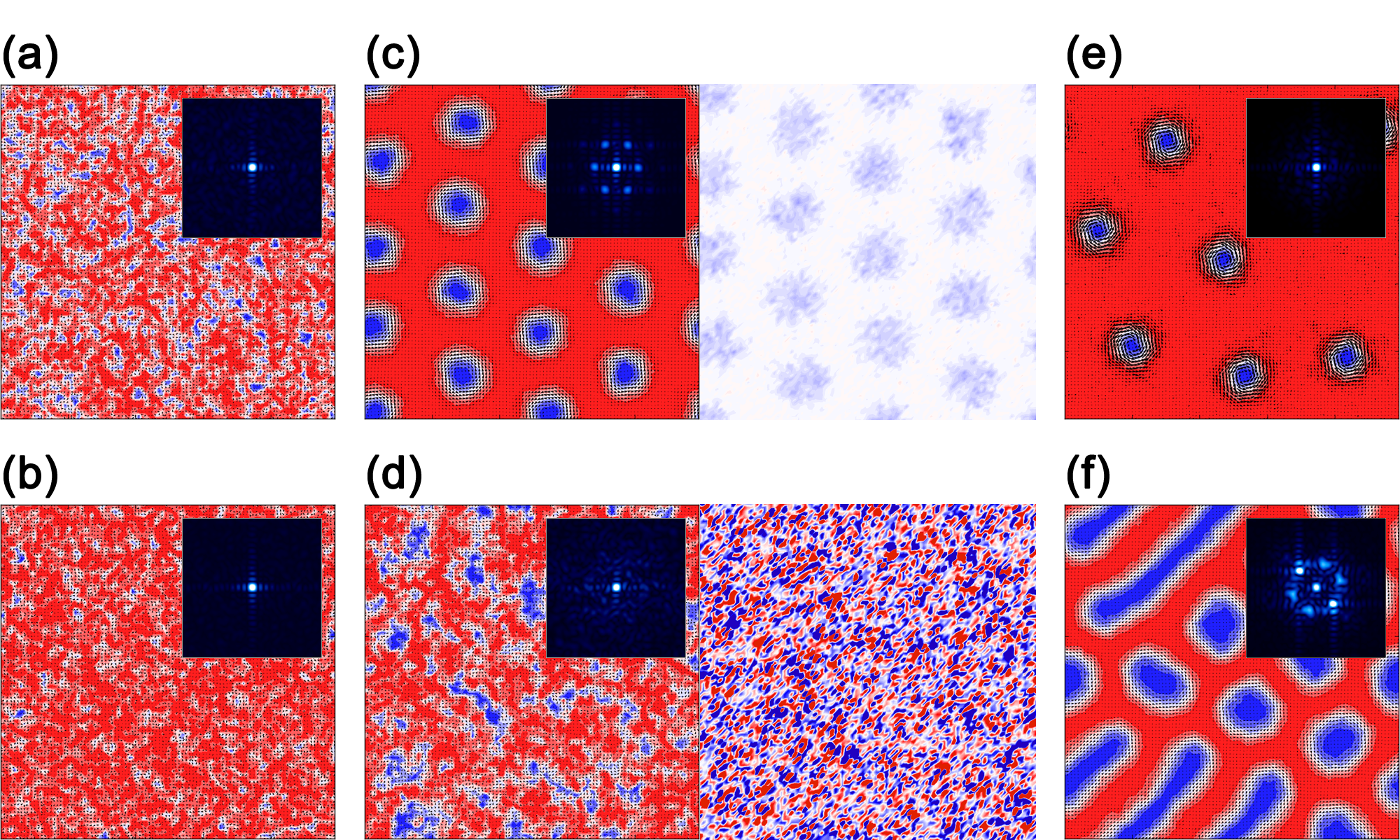} \caption{Snapshots and corresponding reciprocal space plots by Fast Fourier
Transform (FFT) at points on the phase diagram shown in Fig. 2(a).
(a) $B=0.2J$ and $T=1.02J$ , (b) $B=0.2J$ and $T=0.80J$ , (c)
$B=0.06J$ and $T=0.02J$ , (d) $B=0.06J$ and $T=0.66J$, (e) $B=0.08J$
and $T=0.02J$and (f) $B=0.02J$ and $T=0.02J$. In real space snapshots,
red (blue) contour represent the positive (negative) value of $S_{iz}$
and the arrows represent the directions of in-plane component. For
(c) and (d), the density of TC is also shown at right panel respectively.}
\label{fig:snapshots} 
\end{figure}

This emergent topology at finite temperatures does not correspond
to any ordered phase such as the skyrmion crystal phase (SkX) or meron-helix
composite. Two snapshots of spin states around the ridge were taken,
as shown in Fig. \ref{fig:snapshots}(a) and (b). Location of their
corresponding parameters are labeled by the same letter in the $B$-$T$
phase diagram in Fig. \ref{fig:phase}. At point A to the right of
the ridge, $B=0.2J$, $T=1.02J$, and the total TC is about -12 in
a $100\times100$ lattice. However, the real space image shown in
Fig. \ref{fig:snapshots}(a) is completely random. Fast Fourier transformation
of the image provides only one peak at $\Gamma$ point in the reciprocal
space. This indicates the uniform randomness and absence of any spin
ordering at this point. For point B to the left, where the temperature
$T=0.8J$ is relatively lower, the corresponding real space snapshot
in Fig. \ref{fig:snapshots}(b) shows similar randomness with a single
peak at the $\Gamma$ point of the reciprocal space. Compared to point
A, a higher spin polarization parallel with the field is achieved
here. From zero temperature to points A or B of interest, no phase
transition occurs. The emergence of TC is thus purely a consequence
of the thermal fluctuation.

In contrast, TCs at low field, especially at low temperatures, have
distinct origin. Our Monte Carlo simulation shows that the TC grows
significantly around $T=0.25J$ during the annealing procedure and
remains stable to zero temperature. It is attributed to the formation
of the skyrmion crystal phase. A typical snapshot was taken at point
C with $B=0.06J$ and $T=0.02J$ {[}Fig. \ref{fig:snapshots}(c){]}.
The real space image shows a well aligned skyrmion lattice, and the
reciprocal space shows the hexagonal pattern as expected. At the same
field, if the temperature is elevated to point D, the snapshot in
\ref{fig:snapshots}(d) does not present any ordering, although the
TC remains significant. Densities of the TC for C and D points are
ploted in Fig. \ref{fig:snapshots}(c) and Fig. \ref{fig:snapshots}(d)
for comparison. Non-zero TC emerges only near the skyrmion in the
ordered skyrmion phase, while it is evenly distributed in the high
temperature state. At a relatively higher field at point E {[}Fig.
\ref{fig:snapshots}(e){]}, the skyrmion crystal is melted and sparse
skyrmions are observed. While at a lower field at point F, the transition
from skyrmion crystal phase to the helical phase takes place, and
a meron-helix composite appears at this first order phase transition.
In all these regions at low temperatures, the TC is consistent with
the number of skyrmions in the lattice. Thermal fluctuation induced
TC is suppressed. These low-field low-temperature results are consistent
with previous studies\cite{Yi_2009,Buhrandt_2013}.

As indicated by its scaling-free property, origin of the thermally driven
topology can be understood by a simple physical picture on the atomic
scale. As defined earlier, TC is the summation of solid angles of
all triangles in the lattice. Due to the presence of the DM interaction,
these three spins in each triangle are canted, as shown in Fig. \ref{fig:lattice},
and contribute a solid angle of $\Omega$. If we reverse all three
spins, the new configuration cants an opposite solid angle $\Omega$.
In the absence of the field, these two configurations share the same
energy, as both the Heisenberg and DM interactions are quadratic spin
interactions. These two configurations thus have the same probability
of appearance at any temperature, and the average TC is zero. However,
these two configurations, being time reversal to each other, share
opposite magnetizations. An external magnetic field can thus lift
the degeneracy and induce a net TC after thermal averaging. One needs
to be aware that under large enough field, canting of spin takes place
only when the temperature approaches the Curie temperature, far below
which the polarized state is robust and the average TC is zero. On
the other hand, at very high field, the energy difference induced
by the field is no longer relevant, and average TC decays to zero
as well. This well explains the behavior of TC in Fig. \ref{fig:phase}(b).

We can even convey this physical picture in a relatively quantitative
way. Again, focus on a triangle in the lattice with three spins ${\bf S}_{1}$,
${\bf S}_{2}$ and ${\bf S}_{3}$ on the vertices. Notice that ${\bf S}_{2}$
and ${\bf S}_{3}$ are not a pair of nearest neighbors, so no direct
exchange exists between them in our model. The energy of this triangle
is thus given by 
\begin{eqnarray}
E & = & -J({\bf n}_{1}\cdot{\bf n}_{2}+{\bf n}_{1}\cdot{\bf n}_{3})-D(n_{1y}n_{2z}-n_{1z}n_{2y}\nonumber \\
 &  & +n_{1z}n_{3x}-n_{1x}n_{3z})-B(n_{1z}+n_{2z}+n_{3z})\ .
\end{eqnarray}
In the small canting approximation, TC defined in Eq. (\ref{eq:berg})
is simplified as $Q={\bf n}_{1}\cdot({\bf n}_{2}\times{\bf n}_{3})$.
Thermal average of TC is $\langle Q\rangle=\text{\ensuremath{\frac{1}{\mathcal{Z}}}}\int\prod_{i}d{\bf n}_{i}Q\exp(-\frac{E}{T})$,
where $\mathcal{Z}=\int\prod_{i}d{\bf n}_{i}\exp(-\frac{E}{T})$ is
the partition function. At the high temperature limit, $E/T\ll1$,
we can expand the Boltzmann distribution in terms of polynomials of
$E/T$. As a result, $\langle Q\rangle=\frac{1}{\mathcal{Z}}\int\prod_{i}d{\bf n}_{i}{\bf n_{1}\cdot}({\bf n}_{2}\times{\bf n}_{3})(1-\frac{E}{T}+\frac{1}{2!}(\frac{E}{T})^{2}-\frac{1}{3!}(\frac{E}{T})^{3}+\mathcal{O}[(\frac{E}{T})^{4}])$.
The leading two orders of $E/T$ vanish because one cannot pair up
all ${\bf n}_{i}$ and their components into even powers. The leading
non-zero term is the third order terms of $E/T$, where non-zero terms
are listed in the Supplementary Materials \cite{Supp}. As a result,
the average TC is proportional to $\frac{D^{2}B}{T^{3}}$. That is
reasonable because the TC respects spatial inversion symmetry but
breaks the time reversal symmetry; the former requires TC to be proportional
to $D$ squared, which is spatial inversion odd, while the latter
enforces linear proportionality between TC and $B$, which is time
reversal odd. No lower order term could meet this symmetry requirement.
This scaling is consistent with the numerical simulation. As shown
in the inset of Fig. \ref{fig:phase}(b), the TC is truly proportional
to the field at high temperatures. The relation between TC and temperature
$T$ is examined at various $D$ values {[}Fig. \ref{fig:phase}(c){]}.
A scaling between peak value of TC and $D$ is shown in the inset,
and a perfect quadratic relation between them is identified. This
quadratic relation is persistent all the way to high temperatures.

Up to now, our handwaving argument is based on only one triangle.
A complete analysis is developed in terms of the $CP^{1}$ formalism
of the spin model. In the continuum limit, the Hamiltonian is given
by $H=\int d^{2}r[\frac{\bar{J}}{2}(\partial_{i}{\bf n})(\partial_{i}{\bf n})-\bar{D}{\bf n}\cdot(\nabla\times{\bf n})-\bar{B}n_{z}]$,
where $i=x,\ y$ and $\bar{J}=JS^{2}$, $\bar{D}=\frac{DS^{2}}{a}$,
and $\bar{B}=\frac{BS}{a^{2}}$ with finite value of $S=|{\bf S}|$
recovered. $a$ is the lattice constant. A normalized two-component
complex field ${\bf z}$ is introduced and let $n_{\mu}={\bf z}^{\dagger}\sigma_{\mu}{\bf z}(\mu=x,y,z)$,
where $\sigma$ are Pauli matrices. In this representation, the Hamiltonian
can be written in terms of a $CP^{1}$ doublet field given by 
\begin{equation}
H=\int d^{2}r2\bar{J}[|(\partial_{i}-i\alpha_{i}+i\kappa\sigma_{i}){\bf z}|^{2}-h{\bf z}^{\dagger}\sigma_{z}{\bf z}],
\end{equation}
where $\kappa=\frac{\bar{D}}{2\bar{J}}$ and $h=\frac{\bar{B}}{2\bar{J}}$\cite{CP1-skyrmion}.
$\alpha_{i}=-\frac{i}{2}[{\bf z}^{\dagger}\partial_{i}{\bf z}-(\partial_{i}{\bf z}^{\dagger}){\bf z}]$
is the emergent U(1) gauge field, whose total flux is nothing but
the topological charge defined in Eq. (\ref{eq:TQ}): 
\begin{equation}
Q=\frac{1}{4\pi}\int d^{2}r(\nabla\times\boldsymbol{\alpha})_{z}\ .
\end{equation}
Due to the ${\bf z}$-dependence of $\boldsymbol{\alpha}$, the Hamiltonian
has quartic terms of ${\bf z}$, so the integration over ${\bf z}$
cannot be performed straightforwardly in the partition function $\mathcal{Z}=\int\mathcal{D}{\bf z}^{\dagger}\mathcal{D}{\bf z}\exp(-H/T)$.
Therefore, we rescale the field ${\bf z}\rightarrow\sqrt{\frac{2J}{T}}{\bf z}$
, $\lambda\rightarrow\frac{1}{2}\frac{T}{\bar{J}}\lambda$, define
$f=\frac{T}{\bar{J}}$, and perform the Hubbard-Stratonovich transformation\cite{Auerbach,Supp},
ending up with the partition function: 
\begin{eqnarray}
\mathcal{Z} & = & \int\mathcal{D}{\bf z}^{\dagger}\mathcal{D}{\bf z}\mathcal{D}\alpha_{i}\mathcal{D}\lambda\exp\{-[|(\partial_{i}-i\alpha_{i}+i\kappa\sigma_{i}){\bf z}|^{2}\nonumber \\
 &  & -h{\bf z}^{\dagger}\sigma_{z}{\bf z}+i\lambda({\bf z}^{\dagger}{\bf z}-\frac{2}{f})]\}\label{eq:cp1}
\end{eqnarray}
in which ${\bf z}$ and $\boldsymbol{\alpha}$ are now two independent
dynamical variables. A Lagrange multiplier field $\lambda$
is introduced to enforce the normalization of ${\bf z}$.

\begin{figure}
\includegraphics[scale=0.4]{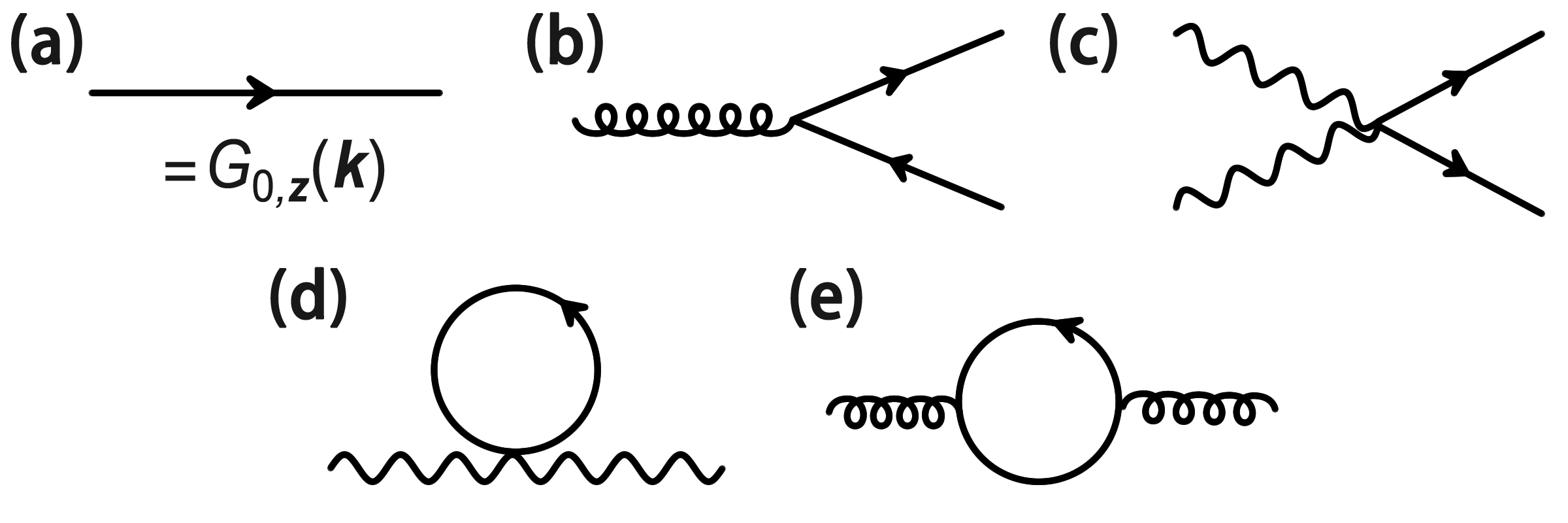} \caption{Feynman rules and diagrams with various integral paths. See details
in the text.}
\label{fig:feynman} 
\end{figure}

The basic idea in what follows is to integrate out the ${\bf z}$
field, and get an effective theory in terms of the gauge field $\boldsymbol{\alpha}$.
The gauge invariance requirement gives rise to only two possible terms
up to the second order of $\boldsymbol{\alpha}$ in the effective
action. One is $b^{2}$ with $b=(\nabla\times\boldsymbol{\alpha})_{z}$
the topological charge density, and the other is $hb$. A saddle point
solution of $b$ thus gives the average value of the TC density proportional
to the field $h$, consistent with the discussions above. To work
it out, a perturbation approach is employed\cite{Oleg,Sachdev}. In momentum space, the
unperturbed part of the action in Eq. (\ref{eq:cp1}) is 
\begin{equation}
S_{0}=L^{2}\int\frac{d^{2}k}{(2\pi)^{2}}{\bf z}_{k}^{\dagger}(k^{2}+m_{0}^{2}-2\kappa k_{i}\sigma_{i}){\bf z}_{k},
\end{equation}
where $L^{2}$ is the area of the 2D film we considered. The corresponding
Feynman diagram is shown in Fig. \ref{fig:feynman}(a). The mass $m_{0}^{2}=i\lambda+2\kappa^{2}$
is determined by the saddle point approximation. Denote the partition
function in Eq. (\ref{eq:cp1}) by $\mathcal{Z}=\int\mathcal{D}\alpha_{i}\mathcal{D}\lambda\exp(-S_{eff}[\alpha_{i},\lambda])$.
A uniform saddle point solution $i\langle\lambda\rangle=\bar{\lambda}$
and $\langle\alpha_{i}\rangle=0$ solves $\delta S_{eff}/\delta\bar{\lambda}=0$,
and we finally get $\log\frac{\Lambda^{2}}{m_{0}^{2}}\approx\frac{4\pi}{f}$\ ,
where $\Lambda\sim\frac{1}{a}$ is the ultraviolet cutoff in Pauli-Villars
regulation scheme\cite{Supp}.

The perturbative part of the action in Eq. (\ref{eq:cp1}) is divided
into two terms 
\begin{eqnarray}
S_{i1} & = & L^{4}\int\frac{d^{2}kd^{2}q}{(2\pi)^{4}}{\bf z}_{k+\frac{q}{2}}^{\dagger}(-2k_{i}\alpha_{i,\ q}-2\kappa\alpha_{i,\ q}\sigma_{i}\nonumber \\
 &  & -h_{q}\sigma_{3}){\bf z}_{k-\frac{q}{2}},\\
S_{i2} & = & L^{6}\int\frac{d^{2}kd^{2}qd^{2}q}{(2\pi)^{4}}{\bf z}_{k}^{\dagger}{\bf z}_{q}\alpha_{i,p}\alpha_{i,k-q-p}.
\end{eqnarray}
The Feynman diagram Fig. \ref{fig:feynman}(b) corresponds to $S_{i1}$,
where the spring line represents the part $-2(k_{i}\alpha_{i,\ q}+\kappa\alpha_{i,\ q}\sigma_{i}+\frac{h}{2}\sigma_{3})$
in the three-point vertex. Fig. \ref{fig:feynman}(c) is four point
interaction in $S_{i2}$. The tilde line represents the pure emergent
gauge field $\alpha_{i}$ of the four-point vertex. The first order perturbation
from $S_{i2}$, shown by the diagram Fig. \ref{fig:feynman}(d), contributes
to a term $S_{d}=\frac{L^{4}}{2\pi}\log\frac{\Lambda^{2}}{m_{0}^{2}}\int\frac{d^{2}q}{(2\pi)^{2}}\alpha_{i}^{2}$.
We don't consider the quadratic term of $\alpha_{i}$ at $\kappa^{2}$
and higher order because $\frac{\kappa^{2}}{\Lambda^{2}}\ll1$. In
contrast, the first order perturbation of $S_{i1}$ is a vanishing
tadpole diagram. The lowest contribution is the second order perturbation
$S_{e}$ depicted in Fig. \ref{fig:feynman}(e). $S_{e}$ can be split
into two parts, $S_{e1}$ correspond to the $b^{2}$ term and $S_{e2}$
which includes the $hb$ term. Combining $S_{d}$ with $S_{e1}$,
we get the gauge invariant term $b^{2}$ as expected: 
\begin{equation}
S_{b^{2}}=S_{d}+S_{e1}=\frac{L^{^{4}}}{\pi}\int\frac{d^{2}q}{(2\pi)^{2}}b_{-q}[\frac{\exp(\frac{4\pi}{f})}{12\Lambda^{2}}+\mathcal{O}(q^{2})]b_{q}.
\end{equation}

The expected $hb$ term also arises from the second order perturbation.
The leading term of $hb$ in $S_{e2}$ is
\begin{equation}
S_{hb}=L^{4}\int\frac{d^{2}kd^{2}q}{(2\pi)^{4}}\frac{8\kappa^{2}(\frac{q^{2}}{4}+m_{0}^{2})h_{-q}b_{q}}{[(k+\frac{q}{2})^{2}+m_{0}^{2}]^{2}[(k-\frac{q}{2})^{2}+m_{0}^{2}]^{2}}
\end{equation}
and the effective action is therefore $S_{eff}=S_{b^{2}}+S_{hb}$.
Solving the saddle point of the field $b$, we obtain
\begin{equation}
\bar{b}=-\frac{12\kappa^{2}h}{\Lambda^{2}}\sinh^{2}(\frac{2\pi}{f})[1-\exp(-\frac{12\pi}{f})].
\end{equation}
The thermal average of the TC at the high temperature limit is 
\begin{equation}
\langle Q\rangle\approx-\frac{18\pi^{2}L^{2}BS^{5}}{T^{3}}(\frac{D}{a})^{2}[1-\frac{6\pi JS^{2}}{T}+\mathcal{O}(\frac{1}{T^{2}})]\label{eq:Qcp1}\ ,
\end{equation}
where $\frac{D}{a}$ is the DM interaction in the continuum limit.
This result matches well with the simple argument based on one triangle.
Actually, if we further proceed to the fourth order of $E/T$ in the
single triangle argument, a term proportional to $JD^{2}B/T^{4}$
is present, but its sign is opposite to the $1/T^{3}$ term.
The emergent topology at finite temperature can be thus well explained
by this effective theory of the emergent gauge field.

In conclusion, we have discovered thermally driven topology in 2D chiral
magnets. A significant upturn of TC was observed outside the skyrmion
crystal phase. This phenomena is well understood by both analyzing
thermal fluctuations in the atomic scales and field theoretical approach
based on $CP^{1}$ formalism. As has been extensively studied in the
skyrmion physics, non-zero TC would lead to the topological Hall effect,
which was observed in the skyrmion crystal phase only\cite{Neubauer_2009,Huang2012,Li_2013}.
The discrepancy between the topological Hall signal and distribution
of the TC observed in this work is due to the itinerant nature of
the magnetism in most chiral magnets under investigation. Close to
or above the Curie temperature, the local magnetic moment in these
magnets is significantly reduced so that our simulation based on constant
local magnetic moment does not apply. Only in insulating magnets such
as Cu$_{2}$OSeO$_{3}$\cite{Seki2012}, local magnetic moments are
persistent at elevated temperatures, and our discovery would apply.
On the other hand, the thermal Hall effect related to the to the magnon
deflection by TC has been addressed in frustrated magnets\cite{Onose_2010,Hirschberger2015,Lee2015}
and chiral magnets \cite{Mochizuki2014,Iwasaki}. We therefore predict
the thermal Hall effect of insulating chiral magnets, in which local
magnetic moments are persistent at high fields and temperatures. Actually the phenomenon of thermally driven topology can be even generalized to ferroelectrics\cite{add001}, and we would expect rich experimental observations will come out in the future.

Upon finishing this work, we noticed that similar behavior of the
topological charge was recently studied by  Levente R$\acute{{\rm o}}$zsa \textit{et al.}\cite{R_zsa_2016}
and Mohit Randeria. Both of them studied the skyrmion crystal phase
rather than the high field case we are emphasizing in this work. We also noticed a recent work\cite{add002} which addressed the similar phenomenon in terms of skyrmion-antiskyrmion formations. We
acknowledge initial discussions with Jung Hoon Han, Oleg Tchernyshyov
and Di Xiao. This work was supported by the U.S. Department of Energy
(DOE), Office of Science, Basic Energy Sciences (BES) under Award
No. DE-SC0016424.

\pagebreak
\widetext
\begin{center}
\textbf{\large Supplementary Materials for "Thermally driven topology in chiral magnets"}
\end{center}

\subsection*{Quantitative analysis on a single triangle in the lattice }
subsection*{Quantitative analysis on a single triangle in the lattice }

The energy of a single triangle consisting of spins ${\bf S}_{1}$, ${\bf S}_{2}$
and ${\bf S}_{3}$ as shown in Fig.1 of the main text, is given by 
\begin{equation}
E=-J({\bf n}_{1}{\bf \centerdot}{\bf n}_{2}+{\bf n}_{1}\centerdot{\bf n}_{3})-D(n_{1y}n_{2z}-n_{1z}n_{2y}+n_{1z}n_{3x}-n_{3z}n_{1x})-B(n_{1z}+n_{2z}+n_{3z})\ ,
\end{equation}
where ${\bf n}_{i}(i=1,2,3)$ are the normalized spin vectors.  Topological
charge (TC) is defined as $Q={\bf n}_{1}\centerdot({\bf n}_{2}\times{\bf n}_{3})$, and has a
thermal average $\langle Q\rangle=\frac{1}{\mathcal{Z}}\int\prod_{i}dn_{i}Q\exp(-\frac{E}{T})$,
where $\mathcal{Z}=\int\prod_{i}dn_{i}\exp(-\frac{E}{T})$ is the
partition function. In the high temperature limit, $E\ll T$, we can expand
the thermal average of the TC as $\langle Q\rangle=\frac{1}{\mathcal{Z}}\int\prod_{i}dn_{i}Q[1-\frac{E}{T}+\frac{1}{2}\frac{E^{2}}{T^{2}}-\frac{1}{3!}\frac{E^{3}}{T^{3}}+\frac{1}{4!}\frac{E^{4}}{T^{4}}+...]$.
Using the parameterization of normalized spin vectors, ${\bf n}_{i}=(\sin\theta_{i}\cos\phi_{i},\ \sin\theta_{i}\sin\phi_{i},\cos\theta_{i})$, the replacement $\int\prod_{i}dn_{i}=\int\prod_{i}d\Omega_{i}$ can be made,
where $\int d\Omega_{i}=\int_{0}^{2\pi}d\phi_{i}\int_{0}^{\pi}\sin\theta_{i}d\theta_{i}$
. Nonzero terms in the polynomial expansion of $E/T$ must include
${\bf n_{i}}$ and their components with even powers, of which, the leading
two orders vanish, since at these orders,  ${\bf n}_{i}$
cannot be paired with their components into even powers. All the nonzero contributions
of order $\frac{1}{T^{3}}$ and $\frac{1}{T^{4}}$ are given by:

\begin{eqnarray}
\mathcal{O}(\frac{1}{T^{3}}):\ &-\frac{(-D)^{2}(-B)}{T^{3}}n_{1y}n_{2z}n_{1z}n_{3x}n_{1z}(n_{1y}n_{2z}n_{3x})= \frac{D^{2}B}{T^{3}}(n_{1y}n_{2z}n_{3x})^{2}n_{1z}^{2},\nonumber \\
&-\frac{(-D)^{2}(-B)}{T^{3}}n_{1z}n_{2y}n_{3z}n_{1x}n_{1z}(n_{1x}n_{2y}n_{3z}) = \frac{D^{2}B}{T^{3}}(n_{1x}n_{2y}n_{3z})^{2}n_{1z}^{2},\nonumber \\
&-\frac{(-D)^{2}(-B)}{T^{3}}(-n_{1z}n_{2y})n_{1z}n_{3x}n_{1z}(-n_{1z}n_{2y}n_{3x}) = \frac{D^{2}B}{T^{3}}(n_{1z}n_{2y}n_{3x})^{2}n_{1z}^{2}.\\
\mathcal{O}(\frac{1}{T^{4}}):\ &\frac{(-D)^{2}(-J)(-B)}{T^{4}}n_{1y}n_{2z}n_{1z}n_{3x}n_{1z}n_{2z}n_{2z}(n_{1y}n_{2z}n_{3x}) = \frac{D^{2}JB}{T^{4}}(n_{1y}n_{2z}n_{3x})^{2}n_{1z}^{2}n_{2z}^{2},\nonumber \\
&\frac{(-D)^{2}(-J)(-B)}{T^{4}}n_{1y}n_{2z}n_{1z}n_{3x}n_{1y}n_{2y}n_{2z}(-n_{1z}n_{2y}n_{3x}) = -\frac{D^{2}JB}{T^{4}}(n_{1z}n_{2y}n_{3x})^{2}n_{1y}^{2}n_{2z}^{2},\nonumber \\
&\frac{(-D)^{2}(-J)(-B)}{T^{4}}n_{1z}n_{2y}n_{3z}n_{1x}n_{1z}n_{3z}n_{1z}(n_{1x}n_{2y}n_{3z}) = \frac{D^{2}JB}{T^{4}}(n_{1x}n_{2y}n_{3z})^{2}n_{1z}^{2}n_{3z}^{2},\nonumber \\
&\frac{(-D)^{2}(-J)(-B)}{T^{4}}n_{1z}n_{2y}n_{3z}n_{1x}n_{1x}n_{3x}n_{3z}(-n_{1z}n_{2y}n_{3x}) = -\frac{D^{2}JB}{T^{4}}(n_{1z}n_{2y}n_{3x})^{2}n_{1x}^{2}n_{3z}^{2},\nonumber \\
&\frac{(-D)^{2}(-J)(-B)}{T^{4}}n_{1y}n_{2z}(-n_{3z}n_{1x})n_{1y}n_{2y}n_{2z}(n_{1x}n_{2y}n_{3z}) = -\frac{D^{2}JB}{T^{4}}(n_{1x}n_{2y}n_{3z})^{2}n_{1y}^{2}n_{2z}^{2},\nonumber \\
&\frac{(-D)^{2}(-J)(-B)}{T^{4}}n{}_{1y}n_{2z}(-n_{3z}n_{1x})n_{1x}n_{3x}n_{3z}(n_{1y}n_{2z}n_{3x}) = -\frac{D^{2}JB}{T^{4}}(n_{1y}n_{2z}n_{3x})^{2}n_{3z}^{2}n_{1x}^{2},\nonumber \\
&\frac{(-D)^{2}(-J)(-B)}{T^{4}}(-n_{1z}n_{2y})n_{1z}n_{3x}n_{1z}n_{2z}n_{2z}(-n_{1z}n_{2y}n_{3x}) = \frac{D^{2}JB}{T^{4}}(n_{1z}n_{2y}n_{3x})^{2}n_{1z}^{2}n_{2z}^{2},\nonumber \\
&\frac{(-D)^{2}(-J)(-B)}{T^{4}}(-n_{1z}n_{2y})n_{1z}n_{3x}n_{1z}n_{3z}n_{3z}(-n_{1z}n_{2y}n_{3x}) = \frac{D^{2}JB}{T^{4}}(n_{1z}n_{2y}n_{3x})^{2}n_{1z}^{2}n_{3z}^{2},\nonumber \\
&\frac{(-D)^{2}(-J)(-B)}{T^{4}}(-n_{1z}n_{2y})n_{1z}n_{3x}n_{1x}n_{3x}n_{3z}(n_{1x}n_{2y}n_{3z}) = -\frac{D^{2}JB}{T^{4}}(n_{1x}n_{2y}n_{3z})^{2}n_{1z}^{2}n_{3x}^{2}, \nonumber \\
&\frac{(-D)^{2}(-J)(-B)}{T^{4}}(-n_{1z}n_{2y})n_{1z}n_{3x}n_{1y}n_{2y}n_{2z}(n_{1y}n_{2z}n_{3x}) = -\frac{D^{2}JB}{T^{4}}(n_{1y}n_{2z}n_{3x})^{2}n_{1z}^{2}n_{2y}^{2}.
\end{eqnarray}
The $+/-$ signs are determined by the number of times ${\bf n}_{1},\ {\bf n}_{2}$ 
and ${\bf n}_{3}$ are exchanged. After adding together the terms of each order, it is found that the contribution to the TC from order $\frac{1}{T^{3}}$ is positive, and order $\frac{1}{T^{4}}$ is negative. The flipping triangle with
${\bf S}'_{1}$, ${\bf S}'_{2}$ and ${\bf S'}_{3}$ has the opposite
contributions to the TC. 

\subsection*{The energy in the continuum limit and the Hubbard-Stratonovich transformation}

We consider an $N\times N$ square lattice system with a lattice constant $a$, and continuum limit energy

\begin{equation}
H=\int d^{2}r[\frac{\bar{J}}{2}(\partial_{i}{\bf n})(\partial_{i}{\bf n})+\bar{D}{\bf n}\centerdot(\nabla\times{\bf n})-\bar{B}n_{z}],
\end{equation}
where $i=x,\ y$, $\bar{J}=JS^{2}$, $\bar{D}=\frac{DS^{2}}{a}$, $\bar{B}=\frac{g\mu_{B}H'S}{a^{2}}=\frac{BS}{a^{2}}$ and $B=g\mu_{B}H'$. $J$ is the Heisenberg interaction, $D$ is the DM interaction and $H'$ is the applied magnetic field along the $z$ axis. In the $CP^{1}$ model, $n_{\mu}={\bf z}^{\dagger}\sigma_{\mu}{\bf z}(\mu=x,y,z)$ and ${\bf z}$
is a two component spinor. The energy density is
\begin{equation}
\mathcal{\mathcal{H}}=2\bar{J}(\partial_{i}{\bf z})^{\dagger}(\partial_{i}{\bf z})-4\bar{J}A_{i}^{2}-2\bar{D}{\bf n}\centerdot{\bf A}-i\bar{D}{\bf z}^{\dagger}(\sigma\centerdot{\bf A}){\bf z}+i\bar{D}(\nabla{\bf z}^{\dagger})\centerdot\sigma{\bf z}-\bar{B}{\bf z}^{\dagger}\sigma_{z}{\bf z},\label{eq:den}
\end{equation}
where $A_{i}=-\frac{i}{2}[{\bf z}^{\dagger}(\partial_{i}{\bf z})-(\partial_{i}{\bf z})^{\dagger}{\bf z}]$.
We perform the Hubbard-Stratonovich transformation to decouple the quartic terms of field ${\bf z}$. In the $CP^{1}$ representation, the partition function is 
\begin{equation}
\mathcal{Z}=\int\mathcal{D}\alpha_{i}\mathcal{D}{\bf z^{\dagger}\mathcal{D}z}\exp\{-\frac{1}{T}\int d^{2}r[2\bar{J}|(\partial_{i}-i\alpha_{i}+i\kappa\sigma_{i}){\bf z}|^{2}-\bar{B}{\bf z}^{\dagger}\sigma_{z}{\bf z}]\}\delta({\bf z}^{\dagger}{\bf z}-1)\},
\end{equation}
where $\bm{\alpha}$ is the emergent $U(1)$ gauge field mentioned in the main text and $\kappa=\frac{\bar{D}}{2\bar{J}}$. We can transform the partition function with quadratic terms of $\alpha_{i}$,
\begin{eqnarray}
\mathcal{Z} & = & \int\mathcal{D}\alpha_{i}\mathcal{D}{\bf z}^{\dagger}\mathcal{D}{\bf z}\nonumber \\
 &  & \times\exp\{-\frac{1}{T}\int d^{2}r[2\bar{J}(\alpha_{i}-A_{i}-\kappa{\bf z}^{\dagger}\sigma_{i}{\bf z})^{2}+4\bar{J}\kappa^{2}-\bar{B}{\bf z}^{\dagger}\sigma_{z}{\bf z}-2\bar{D}{\bf z}^{\dagger}\sigma_{i}{\bf z}A_{i}\nonumber \\
 &  & +2\bar{J}[(\partial_{i}{\bf z})^{\dagger}(\partial_{i}{\bf z})-A_{i}^{2}]+i\bar{D}(\partial_{i}{\bf z}^{\dagger}\sigma_{i}{\bf z}-{\bf z}^{\dagger}\sigma_{i}\partial_{i}{\bf z})]\}\delta({\bf z}^{\dagger}{\bf z}-1).
\end{eqnarray}
After integrating the fields $\alpha_{i}$ out, the partition function becomes 

\begin{eqnarray}
\mathcal{Z} & = & \mathcal{C}\int\mathcal{D}{\bf z^{\dagger}\mathcal{D}z}\exp\{-\frac{1}{T}\int d^{2}r[2\bar{J}[(\partial_{i}{\bf z})^{\dagger}(\partial_{i}{\bf z})-A_{\mu}^{2}-\bar{B}{\bf z}^{\dagger}\sigma_{z}{\bf z}\nonumber \\
 &  & +i\bar{D}(\partial_{\mu}{\bf z}^{\dagger}\sigma_{\mu}{\bf z}-{\bf z}^{\dagger}\sigma_{\mu}\partial_{\mu}{\bf z})-2\bar{D}{\bf z}^{\dagger}\sigma_{\mu}{\bf z}A_{\mu}+4\bar{J}\kappa^{2}]\}\delta({\bf z}^{\dagger}{\bf z}-1),
\end{eqnarray}
where  $\mathcal{C}$ is a constant  from the integration. The effective Hamiltonian is 
\begin{eqnarray}
H_{eff} & = & \int d^{2}r[2\bar{J}[(\partial_{\mu}{\bf z})^{\dagger}(\partial_{\mu}{\bf z})-A_{\mu}^{2}]+i\bar{D}(\partial_{\mu}{\bf z}^{\dagger}\sigma_{\mu}{\bf z}-{\bf z}^{\dagger}\sigma_{\mu}\partial_{\mu}{\bf z})\nonumber \\
 &  & -2\bar{D}{\bf z}^{\dagger}\sigma_{\mu}{\bf z}A_{\mu}-\bar{B}{\bf z}^{\dagger}\sigma_{z}{\bf z}
\end{eqnarray}
which is as same as Eqn.(\ref{eq:den}).

\subsection*{Mean field approximation of the constraint field}

We extend the $CP^{1}$ model to the $CP^{\mathcal{N}-1}$ model in
which the field ${\bf z}$ has $\mathcal{N}$ flavors and $|{\bf z}^{\dagger}{\bf z}|=\frac{\mathcal{N}}{2}$.
The fields can be rescaled as ${\bf z}\rightarrow\sqrt{\frac{2\bar{J}}{T}}{\bf z}$
and define $h=\frac{\bar{B}}{2\bar{J}}$, $f=\frac{T}{\bar{J}}$ 
and $\lambda\rightarrow\frac{f}{2}\lambda$. The partition function
transforms into

\begin{equation}
\mathcal{Z}=\int\mathcal{D}{\bf z^{\dagger}\mathcal{D}z\mathcal{D}}\alpha_{i}\mathcal{D}\lambda\exp\{-[|(\partial_{i}-i\alpha_{i}+i\kappa\sigma_{i}){\bf z}|^{2}-h{\bf z}^{\dagger}\sigma_{z}{\bf z}-i\lambda({\bf z}^{\dagger}{\bf z}-\frac{\mathcal{N}}{f})\}.
\end{equation}
After integrating out the field ${\bf z}$, the partition function
has the form $\mathcal{Z}=\int\mathcal{D}\alpha_{i}\mathcal{D}\lambda\exp(-S_{eff}[\alpha_{i},\lambda])$, where

\begin{equation}
S_{eff}[\alpha_{i,}\lambda]=C'+Tr\log[-(\partial_{i}-i\alpha_{i}+i\kappa\sigma_{i})^{2}+h\sigma_{z}+i\lambda]-\frac{\mathcal{N}i}{f}\int d^{2}r\lambda
\end{equation}
and $C'$ is a constant. When we consider the $\mathcal{N}\rightarrow$$\infty$ limit, the effective action can be approximated by the quadratic fluctuation
around the saddle point. The saddle point is $i\langle\lambda\rangle=\bar{\lambda},\ \langle\alpha_{i}\rangle=0$.
We can ignore the Zeeman coupling term in the large $\mathcal{N}$ limit with a finite temperature, since $h\ll|\frac{\mathcal{N}\bar{\lambda}}{f}|$ when $\mathcal{N}\rightarrow\infty$. The effective action around
saddle point in momentum space is 

\begin{equation}
S_{eff}[0,\ \bar{\lambda}]=C''+\sum_{k}\log[(k^{2}+\bar{\lambda}+2\kappa^{2})^{2}-4\kappa^{2}k^{2}]-\frac{\mathcal{N}L^{2}\bar{\lambda}}{f},
\end{equation}
where $L^{2}=N^{2}a^{2}$ is the area of the space. Here, we use the
relationships  $\sigma_{3}\sigma_{i}\sigma_{3}=-\sigma_{i}$ and
$Tr\log(ABC)=Tr\log(CAB)$ to work out the trace. By replacing $\sum_{k}$by
$L^{2}\int\frac{d^{2}k}{(2\pi)^{2}}$, we can obtain 

\begin{equation}
\frac{1}{2}\int_{-\Lambda}^{\Lambda}\frac{d^{2}k}{(2\pi)^{2}}\frac{2(k^{2}+\bar{\lambda}+2\kappa^{2})}{(k^{2}+\bar{\lambda}+2\kappa^{2})^{2}-4\kappa^{2}k^{2}}-\frac{\mathcal{N}}{f}=0\, .
\end{equation}
If we consider a finite size system, the momentum in the integral has
a cutoff $\Lambda\sim\frac{1}{a}$. Based on the assumption that $\kappa^{2}\ll\bar{\lambda}<\Lambda^{2}$, the saddle point equation transforms into 

\begin{equation}
\log\frac{\Lambda^{2}+\bar{\lambda}+2\kappa^{2}}{\bar{\lambda}+2\kappa^{2}}+\frac{2\kappa}{\sqrt{\bar{\lambda}+\kappa^{2}}}\arctan\frac{\kappa}{\sqrt{\bar{\lambda}+\kappa^{2}}}\log\approx\frac{2\pi\mathcal{N}}{f} \, ,
\end{equation}
where the second term on the left side can be neglected. Turning back to the $CP^{1}(\mathcal{N}=2)$ model, the solution has a simple form 

\begin{equation}
\log\frac{\Lambda^{2}+m_{0}^{2}}{m_{0}^{2}}\approx\frac{4\pi}{f},\label{eq:Cut-Off}
\end{equation}
where $m_{0}^{2}=\bar{\lambda}+2\kappa^{2}$. This is the momentum
cutoff scheme\cite{Cutoff}. We can also employ the Pauli-Villars regularization,
which protects the gauge symmetry and translational symmetry. We
integral over $k$ from $-\infty$ to $\infty$, and replace $\int_{-\infty}^{\infty}\frac{d^{2}k}{(2\pi)^{2}}\frac{2(k^{2}+\bar{\lambda}+2\kappa^{2})}{(k^{2}+\bar{\lambda}+2\kappa^{2})^{2}-4\kappa^{2}k^{2}}$
by $\frac{1}{2}\int_{-\infty}^{\infty}\frac{d^{2}k}{(2\pi)^{2}}\frac{2(k^{2}+\bar{\lambda}+2\kappa^{2})}{(k^{2}+\bar{\lambda}+2\kappa^{2})^{2}-4\kappa^{2}k^{2}}-\int_{-\infty}^{\infty}\frac{d^{2}k}{(2\pi)^{2}}\frac{1}{k^{2}+\Lambda_{PV}^{2}}$,
where $\Lambda_{PV}$ is the cutoff. The solution in Pauli-Villars
regularization is 

\begin{equation}
\log\frac{\Lambda_{PV}^{2}}{m_{0}^{2}}\approx\frac{4\pi}{f} \, .\label{eq:PV}
\end{equation}
In the momentum cutoff scheme, there is no need to assume $m_{0}^{2}\ll\Lambda^{2}$, so we can use this model when $m_{0}$ is comparable with the cut-off $\Lambda$.  In the very low temperature region $(f\ll1)$, we also
get $m_{0}^{2}\ll\Lambda^{2}$; therefore,  at low
temperature, $\log\frac{\Lambda^{2}}{m_{0}^{2}}\approx\frac{4\pi}{f}$
works in both schemes.

\subsection*{Perturbative Calculation}

In momentum space, the action can be split into the unperturbed part, $S_{0}$,  and perturbed parts, $S_{i1}$, and $S_{i2}$

\begin{eqnarray}
S_{0} & = & \sum_{k}{\bf z}_{k}(k^{2}+m_{0}^{2}-2\kappa k_{i}\sigma_{i}){\bf z}_{k},\\
S_{i1} & = & -\sum_{k,q}{\bf z}_{k+\frac{q}{2}}^{\dagger}(2k_{i}\alpha_{i,\ q}+2\kappa\alpha_{i,\ q}\sigma_{i}+h_{q}\sigma_{z}){\bf z}_{k-\frac{q}{2}},\\
S_{i2} & = & \sum_{k,q,p}{\bf z}_{k}^{\dagger}{\bf {\bf z}}_{q}\alpha_{i,p}\alpha_{i,k-p-q},
\end{eqnarray}
where we treat the applied field $h$ as a local field. The Green's
function of the field ${\bf z}$ is $G_{0,z}(k)=\langle{\bf z}_{k}{\bf z}_{k}^{\dagger}\rangle=\frac{1}{k^{2}+m_{0}^{2}-2\kappa k_{i}\sigma_{i}}$. As discussed in the main text, Fig.4(b) corresponds to the interaction
described by $S_{i1}$. In the perturbative calculation, we replace
$\sum_{k}$ by $L^{2}\int\frac{d^{2}k}{(2\pi)^{2}}$, and the action described by the process in Fig.4(d) is

\begin{eqnarray}
S_{d} & = & L^{4}Tr\int\frac{d^{2}kd^{2}q}{(2\pi)^{4}}\frac{\alpha_{i,q}\alpha_{i,-q}}{k^{2}+m_{0}^{2}-2\kappa k_{i}\sigma_{i}}\nonumber\\
& = &L^{4}\int\frac{d^{2}kd^{2}q}{(2\pi)^{4}}[\frac{2\alpha_{i,q}\alpha_{i,q}}{k^{2}+m_{0}^{2}}+\frac{4\kappa^{2}k^{2}\alpha_{i,q}^{2}}{(k^{2}+m_{0}^{2})^{4}}+\mathcal{O}(\kappa^{4})]\, ,
\end{eqnarray}

where the terms of order $\kappa^{2}$ and higher can be neglected, since $\frac{\kappa^{2}}{\Lambda^{2}}\ll1$. Pauli-Villars regularization is applied to the divergent integral,

\begin{eqnarray}
\int\frac{d^{2}k}{(2\pi)^{2}}\frac{1}{k^{2}+m_{0}^{2}} & \rightarrow & \int\frac{d^{2}k}{(2\pi)^{2}}(\frac{1}{k^{2}+m_{0}^{2}}-\frac{1}{k^{2}+\Lambda_{PV}^{2}})\nonumber \\
 & = & \frac{1}{4\pi}\log\frac{\Lambda_{PV}^{2}}{m_{0}^{2}} \, ,
\end{eqnarray}
so that 

\begin{equation}
S_{d}=\frac{L^{4}}{2\pi}\log\frac{\Lambda_{PV}^{2}}{m_{0}^{2}}\int\frac{d^{2}q}{(2\pi)^{2}}\alpha_{i,q}\alpha_{i,-q} \, \, \, \, \, \, .
\end{equation}
The process in Fig4.(e) corresponds to the action 

\begin{eqnarray}
S_{e} & = & -\frac{L^{4}}{2!}\int\frac{d^{2}kd^{2}q}{(2\pi)^{4}}\nonumber \\
 &  & \times\langle2{\bf z}_{k+\frac{q}{2}}^{\dagger}(-k_{i}\alpha_{i,q}-\kappa\alpha_{i,q}\sigma_{i}-\frac{h_{q}}{2}\sigma_{z}){\bf z}{}_{k-\frac{q}{2}}\nonumber \\
 &  & \times2{\bf z}_{k-\frac{q}{2}}^{\dagger}(-k_{j}\alpha_{j,-q}-\kappa\alpha_{j,-q}\sigma_{j}-\frac{h_{-q}}{2}\sigma_{z}){\bf z}_{k+\frac{q}{2}}\rangle\nonumber \\
 & = & -2L^{4}Tr\int\frac{d^{2}kd^{2}q}{(2\pi)^{4}}\frac{1}{(k+\frac{q}{2})^{2}+m_{0}^{2}-2\kappa(k+\frac{q}{2})_{i'}\sigma_{i'}}(k_{i}\alpha_{i,-q}+\kappa\alpha_{i,-q}\sigma_{i}+\frac{h_{-q}}{2}\sigma_{z})\nonumber \\
 &  & \times\frac{1}{(k-\frac{q}{2})^{2}+m_{0}^{2}-2\kappa(k-\frac{q}{2})_{j'}\sigma_{j'}}(k_{j}\alpha_{j,q}+\kappa\alpha_{j,q}\sigma_{j}+\frac{h_{q}}{2}\sigma_{z})\nonumber \\
 & = & -2L^{4}\int\frac{d^{2}kd^{2}q}{(2\pi)^{4}}[\frac{1}{(k+\frac{q}{2})^{2}+m_{0}^{2}}+\frac{2\kappa(k+\frac{q}{2})_{i'}\sigma_{i'}}{[(k+\frac{q}{2})^{2}+m_{0}^{2}]^{2}}+\mathcal{O}(\kappa^{2})]\nonumber \\
 &  & \times(k_{i}\alpha_{i,-q}+\kappa\alpha_{i,-q}\sigma_{i}+\frac{h_{-q}}{2}\sigma_{z})[\frac{1}{(k-\frac{q}{2})^{2}+m_{0}^{2}}\nonumber \\
 &  & +\frac{2\kappa(k-\frac{q}{2})_{j'}\sigma_{j'}}{[(k-\frac{q}{2})^{2}+m_{0}^{2}]^{2}}+\mathcal{O}(\kappa^{2})](k_{j}\alpha_{j,q}+\kappa\alpha_{j,q}\sigma_{j}+\frac{h_{q}}{2}\sigma_{z})\nonumber \\
 & = & -2L^{4}Tr\int\frac{d^{2}kd^{2}q}{(2\pi)^{4}}[\frac{k^{2}\alpha_{q}^{2}+\kappa^{2}\alpha_{q}^{2}+\frac{h_{q}^{2}}{4}}{[(k+\frac{q}{2})^{2}+m_{0}^{2}][(k-\frac{q}{2})^{2}+m_{0}^{2}]}\nonumber \\
 &  & +\frac{(k_{i}\alpha_{i,-q}+\kappa\alpha_{i,-q}\sigma_{i}+\frac{h_{-q}}{2}\sigma_{z})}{(k+\frac{q}{2})^{2}+m_{0}^{2}}\frac{2\kappa(k-\frac{q}{2})_{j'}\sigma_{j'}}{[(k-\frac{q}{2})^{2}+m_{0}^{2}]^{2}}(k_{j}\alpha_{j,q}+\kappa\alpha_{j,q}\sigma_{j}+\frac{h_{q}}{2}\sigma_{z})\nonumber \\
 &  & +\frac{2\kappa(k+\frac{q}{2})_{i'}\sigma_{i'}}{[(k+\frac{q}{2})^{2}+m_{0}^{2}]^{2}}(k_{i}\alpha_{i,-q}+\kappa\alpha_{i,-q}\sigma_{i}+\frac{h_{-q}}{2}\sigma_{z})\frac{(k_{j}\alpha_{j,q}+\kappa\alpha_{j,q}\sigma_{j}+\frac{h_{q}}{2}\sigma_{z})}{(k-\frac{q}{2})^{2}+m_{0}^{2}}\nonumber \\
 &  & +\mathcal{O}(\kappa^{2})]\, .
\end{eqnarray}

\noindent The $\kappa^{2}\alpha_{i}^{2}$ term is neglected  due to the same reason in $S_{b}$, and the $h^{2}$ term is neglected because it
decouples with $\alpha_{i}$. 

We employ Feynman parametrization to
work out the integrals

\begin{equation}
S_{e1}=-4L^{4}\int\frac{d^{2}qd^{2}k}{(2\pi)^{4}}\int_{0}^{1}dx\frac{(k_{i}\alpha_{i})^{2}}{\{x[(k+\frac{q}{2})^{2}+m_{0}^{2}]+(1-x)[(k-\frac{q}{2})^{2}+m_{0}^{2}]\}^{2}}
\end{equation}

with

\begin{eqnarray}
A & = & \int\frac{d^{2}k}{(2\pi)^{2}}\int_{0}^{1}dx\frac{(k_{i}\alpha_{i})^{2}}{[k^{2}+\frac{q^{2}}{4}+m_{0}^{2}+2(x-\frac{1}{2})k\centerdot q]^{2}}\nonumber \\
 & = & \int_{0}^{1}dx\int\frac{d^{2}k}{(2\pi)^{2}}\frac{(k_{i}\alpha_{i})^{2}}{\{[k+(x-\frac{1}{2})q]^{2}+\frac{q^{2}}{4}+m_{0}^{2}-(x-\frac{1}{2})^{2}q^{2}\}^{2}}\nonumber \\
 & = & \int_{0}^{1}dx\int\frac{d^{2}l}{(2\pi)^{2}}\frac{\{[l-(x-\frac{1}{2})q]_{i}\alpha_{i}\}^{2}}{[l^{2}+m_{0}^{2}+x(1-x)q^{2}]^{2}},
\end{eqnarray}
where $l=k+(x-\frac{1}{2})q$ and $\Delta\equiv m_{0}^{2}+x(1-x)q^{2}$.
$A$ is divided into two parts,

\begin{eqnarray}
A & = & A_{1}+A_{2}\nonumber \\
A_{1} & = & \int_{0}^{1}dx\int\frac{d^{2}l}{(2\pi)^{2}}\frac{(l_{i}\alpha_{i})^{2}}{(l^{2}+\Delta)^{2}}\nonumber \\
 & = & \int_{0}^{1}dx\int_{0}^{2\pi}d\theta\int_{0}^{\infty}\frac{ldl}{(2\pi)^{2}}\frac{(l\alpha\cos\theta)^{2}}{(l^{2}+\Delta)^{2}}\nonumber \\
 & = & \frac{1}{4\pi^{2}}\int_{0}^{2\pi}d\theta\cos^{2}\theta\int_{0}^{1}dx\int_{0}^{\infty}\frac{dl^{2}}{2}\frac{l^{2}}{(l^{2}+\Delta)^{2}}\alpha^{2}\nonumber \\
 & = & \frac{1}{8\pi^{2}}\int_{0}^{1}dx\int_{0}^{2\pi}d\theta(\frac{1+\cos2\theta}{2})\int_{0}^{\infty}dl^{2}\frac{l^{2}\alpha^{2}}{(l^{2}+\Delta)^{2}}\nonumber \\
 & = & \frac{1}{8\pi}\int_{0}^{1}dx\int_{0}^{\infty}dl^{2}\frac{l^{2}\alpha^{2}}{(l^{2}+\Delta)^{2}},\\
A_{2} & = & \frac{1}{4\pi}\int_{0}^{1}dx\int\frac{d^{2}l}{(2\pi)^{2}}\frac{(x-\frac{1}{2})^{2}(q_{i}\alpha_{i})^{2}}{(l^{2}+\Delta)^{2}}\nonumber \\
 & = & \frac{1}{4\pi}\int_{0}^{1}dx\int_{0}^{\infty}dl^{2}\frac{(x-\frac{1}{2})^{2}(q_{i}\alpha_{i})^{2}}{(l^{2}+\Delta)^{2}}\nonumber \\
 & = & \frac{1}{4\pi}\int_{0}^{1}dx[-\frac{(x-\frac{1}{2})^{2}(q_{i}\alpha_{i})^{2}}{l^{2}+\Delta}]\bigg|_{l^{2}=0}^{l^{2}=\infty}\nonumber \\
 & = & \frac{1}{4\pi}\int_{0}^{1}dx\frac{(x-\frac{1}{2})^{2}(q_{i}\alpha_{i})^{2}}{m_{0}^{2}+x(1-x)q^{2}}\nonumber \\
 & = & \frac{1}{4\pi}\frac{(q_{i}\alpha_{i})^{2}}{q^{2}}[-1+\frac{1}{2}\sqrt{\frac{q^{2}+4m_{0}^{2}}{q^{2}}}\log(\frac{\sqrt{q^{2}+4m_{0}^{2}}+q}{\sqrt{q^{2}+4m_{0}^{2}-q}})]\, ,
\end{eqnarray}
where the integral in $A_{1}$ is divergent. Pauli-Villars
regularization is again applied to deduct the divergent part

\begin{eqnarray}
A_{1}\rightarrow A'_{1} & = & \frac{1}{8\pi}\int_{0}^{1}dx\int_{0}^{\infty}dl^{2}\{\frac{l^{2}\alpha_{i}^{2}}{[l^{2}+m_{0}^{2}+x(1-x)q^{2}]^{2}}-\frac{l^{2}\alpha_{i}^{2}}{(l^{2}+\Lambda_{PV}^{2})^{2}}\}\nonumber \\
 & = & \frac{1}{8\pi}\int_{0}^{1}dx(\log\frac{\Lambda_{PV}^{2}}{m_{0}^{2}+x(1-x)q^{2}})\alpha_{i}^{2}\nonumber \\
 & = & \frac{1}{8\pi}\int_{0}^{1}dx\{\log\Lambda_{PV}^{2}-\log[m_{0}^{2}+x(1-x)q^{2}]\}\alpha_{i}^{2}\nonumber \\
 & = & \frac{1}{8\pi}[\log\frac{\Lambda_{PV}^{2}}{m_{0}^{2}}+2-\sqrt{\frac{q^{2}+4m_{0}^{2}}{q^{2}}}\log(\frac{\sqrt{q^{2}+4m_{0}^{2}}+|q|}{\sqrt{q^{2}+4m_{0}^{2}}-|q|})]\alpha_{i}^{2}
\end{eqnarray}
where $\alpha_{i}^{2}=\alpha_{i,q}\alpha_{i,-q}$ ,

\begin{eqnarray}
S_{e1} & = & -\frac{L^{4}}{2\pi}\int\frac{d^{2}q}{(2\pi)^{2}}[2-\sqrt{\frac{q^{2}+4m_{0}^{2}}{q^{2}}}\log(\frac{\sqrt{q^{2}+4m_{0}^{2}}+|q|}{\sqrt{q^{2}+4m_{0}^{2}}-|q|})]\nonumber \\
 &  & \times\alpha_{i}(\delta_{ij}-\frac{q_{i}q_{j}}{q^{2}})\alpha_{j}-\frac{1}{2\pi}\log\frac{\Lambda_{PV}^{2}}{m_{0}^{2}}\int\frac{d^{2}q}{(2\pi)^{2}}\alpha_{i}^{2}
\end{eqnarray}
and 

\begin{eqnarray}
S_{\alpha^{2}}=S_{d}+S_{e1} & = & \frac{L^{4}}{\pi}\int\frac{d^{2}q}{(2\pi)^{2}}[\frac{1}{2}\sqrt{\frac{q^{2}+4m_{0}^{2}}{q^{2}}}\log(\frac{\sqrt{q^{2}+4m_{0}^{2}}+|q|}{\sqrt{q^{2}+4m_{0}^{2}}-|q|})-1]\nonumber \\
 &  & \times\alpha_{i}(\delta_{ij}-\frac{q_{i}q_{j}}{q^{2}})\alpha_{j}\, .
\end{eqnarray}
The gauge violation terms in $S_{d}$ and $S_{e1}$ cancel with each
other and we can expand $S_{\alpha^{2}}$ to the order of $q^{2}$ ,

\begin{equation}
S_{\alpha^{2}}=\frac{L^{4}}{\pi}\int\frac{d^{2}q}{(2\pi)^{2}}[\frac{q^{2}}{12m_{0}^{2}}-\frac{(q^{2})^{2}}{120m_{0}^{4}}+\mathcal{O}((q^{2})^{4})]\alpha_{i,\ q}(\delta_{ij}-\frac{q_{i}q_{j}}{q^{2}})\alpha_{j,-q}\, \, \, .
\end{equation}
In the main text, $b=(\nabla\times{\bf \bm{\alpha}})_{z}$ and $S_{\alpha^{2}}$
has the same  form as $S_{b^{2}}$ 

\begin{equation}
S_{b^{2}}=\frac{L^{4}}{\pi}\int\frac{d^{2}q}{(2\pi)^{2}}\frac{b_{q}^{2}}{12m_{0}^{2}}+\mathcal{O}(q^{2}b^{2}),
\end{equation}
where $b_{q}=i\varepsilon_{3ij}q_{i}\alpha_{j,q}$. By using the result in Eqn.(\ref{eq:PV}),

\begin{equation}
S_{b^{2}}=\frac{L^{4}}{\pi}\int\frac{d^{2}q}{(2\pi)^{2}}[\frac{b_{q}^{2}}{12\Lambda_{PV}^{2}}\exp(\frac{4\pi}{f})+\mathcal{O}(q^{2}b^{2})]\, .
\end{equation}
The effective action of the $hb$ term in $S_{e}$ is 
\begin{eqnarray}
S_{e2}=S_{hb} & = & -4L^{4}\int\frac{d^{2}kd^{2}q}{(2\pi)^{4}}[\frac{i\kappa^{2}\varepsilon_{ijz}(k+\frac{q}{2})_{i}\alpha_{j,-q}h_{q}+i\kappa^{2}\varepsilon_{izj}(k+\frac{q}{2})_{i}h_{-q}\alpha_{j,q}}{[(k+\frac{q}{2})^{2}+m_{0}^{2}]^{2}[(k-\frac{q}{2})^{2}+m_{0}^{2}]}\nonumber \\
 &  & +\frac{i\kappa^{2}\varepsilon_{ijz}\alpha_{i,-q}(k-\frac{q}{2})_{j}h_{q}+i\kappa^{2}\varepsilon_{zij}h_{-q}(k-\frac{q}{2})_{i}\alpha_{j,q}}{[(k+\frac{q}{2})^{2}+m_{0}^{2}][(k-\frac{q}{2})^{2}+m_{0}^{2}]^{2}}]\nonumber \\
 & = & 4\kappa^{2}L^{4}\int\frac{d^{2}kd^{2}q}{(2\pi)^{4}}\{\frac{4k\centerdot q(i\varepsilon_{zij})k_{i}(\alpha_{j,-q}h_{q}-\alpha_{j,q}h_{-q})}{[(k+\frac{q}{2})^{2}+m_{0}^{2}]^{2}[(k-\frac{q}{2})^{2}+m_{0}^{2}]^{2}}\nonumber \\
 &  & +\frac{2(k^{2}+\frac{q^{2}}{4}+m_{0}^{2})(b_{q}h_{-q}+b_{-q}h_{q})}{[(k+\frac{q}{2})^{2}+m_{0}^{2}]^{2}[(k-\frac{q}{2})^{2}+m_{0}^{2}]^{2}}\} \, ,
\end{eqnarray}
where $(k\centerdot q)k_{i}$ can be replaced by $\frac{1}{2}k^{2}q_{i}$
in integral\cite{Peskin}. Following the procedure of the Feynman
parametrization used above,

\begin{eqnarray}
S_{hb} & = & 8\kappa^{2}L^{4}\int_{0}^{1}dx\int\frac{d^{2}ld^{2}q}{(2\pi)^{4}}\frac{h_{-q}b_{q}(\frac{q^{2}}{4}+m_{0}^{2})}{(l^{2}+\Delta)^{4}}\nonumber \\
 & = & 8\kappa^{2}L^{4}\int_{0}^{1}dx\int\frac{d^{2}q}{(2\pi)^{2}}\int_{-\Lambda}^{\Lambda}\frac{d^{2}l}{(2\pi)^{2}}\frac{(\frac{q^{2}}{4}+m_{0}^{2})h_{-q}b_{q}}{(l^{2}+\Delta)^{4}} \, \, .
\end{eqnarray}
We consider a finite size system which requires that the momentum has
bounds in the integral. Applying the momentum cutoff scheme  to work
out the integral gives
\begin{equation}
S_{hb}=\frac{2L^{4}\kappa^{2}}{\pi}\int_{0}^{1}dx\int\frac{d^{2}q}{(2\pi)^{2}}[\frac{1}{\Delta^{3}}-\frac{1}{(\Lambda^{2}+\Delta)^{3}}](\frac{q^{2}}{4}+m_{0}^{2})h_{-q}b_{q}\, .
\end{equation}
Expanding the action to the order $q^{2}$, 
\begin{equation}
S_{hb}=\frac{2L^{4}\kappa^{2}}{\pi}\int\frac{d^{2}q}{(2\pi)^{2}}[\frac{(\Lambda^{2}+m_{0}^{2})^{3}-m_{0}^{6}}{m_{0}^{4}(\Lambda^{2}+m_{0}^{2})^{3}}+\mathcal{O}(q^{2})]h_{-q}b_{q}\, \, .
\end{equation}
$S_{b^{2}}$ and $S_{hb}$ are added together, and in position space gives,
\begin{equation}
S_{b^{2}}+S_{hb}=\frac{1}{\pi}\int d^{2}r[\frac{b^{2}(r)}{12\Lambda_{PV}^{2}}\exp(\frac{4\pi}{f})+\frac{(\Lambda^{2}+m_{0}^{2})^{3}-m_{0}^{6}}{2m_{0}^{4}(\Lambda^{2}+m_{0}^{2})^{3}}\kappa^{2}hb(r)+\mathcal{O}(\partial^{2}b)]\, .
\end{equation}
Here, we can simply set $\Lambda_{PV}=\Lambda$. Ignoring the fluctuation
of the $b(r)$, the average value of $b(r)$ is obtained through the
saddle point equation $\frac{\delta(S_{b^{2}}+S_{hb})}{\delta b(r)}=0$,
\begin{equation}
\bar{b}=-\frac{3\kappa^{2}h\Lambda^{2}}{m_{0}^{4}}[1-(\frac{m_{0}^{2}}{\Lambda^{2}+m_{0}^{2}})^{3}]\exp(-\frac{4\pi}{f}) \, .
\end{equation}
The result in Eqn.(\ref{eq:Cut-Off}) is applied to obtain $\bar{b}$
as a function of temperature,
\begin{eqnarray}
\bar{b} & \approx- &\frac{3\kappa^{2}h}{\Lambda^{2}}[\exp(\frac{4\pi}{f})-1]^{2}[1-\exp(-\frac{12\pi}{f})]\exp(-\frac{4\pi}{f})\nonumber \\
 & = &-\frac{12\kappa^{2}h}{\Lambda^{2}}\sinh^{2}(\frac{2\pi}{f})[1-\exp(-\frac{12\pi}{f})].
\end{eqnarray}
At the high temperature limit($\frac{1}{f}\ll1)$, we can expand the $\bar{b}$
by the order of $\frac{1}{f}$,
\begin{equation}
\bar{b}=-\frac{9\kappa^{2}h}{\Lambda^{2}}[(\frac{4\pi}{f})^{3}-\frac{3}{2}(\frac{4\pi}{f})^{4}+\mathcal{O}(\frac{1}{f^{5}})].
\end{equation}
The average of  TC at very high temperature can be approximated as
\begin{eqnarray}
\langle Q\rangle & \approx &\frac{1}{4\pi}\int d^{2}\bar{b}\nonumber\\
 & = & -\frac{N^{2}a^{2}}{4\pi}\frac{9\kappa^{2}h}{\Lambda^{2}}[(\frac{4\pi}{f})^{3}-\frac{3}{2}(\frac{4\pi}{f})^{4}+\mathcal{O}(\frac{1}{f^{5}})].
\end{eqnarray}
By using the parameters in the lattice Hamiltonian $(\kappa=\frac{\bar{D}}{2\bar{J}}=\frac{D}{2Ja},\ h=\frac{\bar{B}}{2\bar{J}}=\frac{B}{2Ja^{2}S},\ f=\frac{T}{\bar{J}}=\frac{T}{JS^{2}})$
with $\Lambda=\frac{1}{a}$, we have 
\begin{eqnarray}
\langle Q\rangle & \approx&-\frac{18\pi^{2}N^{2}D^{2}BS^{5}}{T^{3}}[1-\frac{6\pi JS^{2}}{T}+\mathcal{O}(\frac{1}{T^{2}})]\nonumber\\
 & =&-\frac{18\pi^{2}L^{2}BS^{5}}{T^{3}}(\frac{D}{a})^{2}[1-\frac{6\pi JS^{2}}{T}+\mathcal{O}(\frac{1}{T^{2}})]\label{eq:TC},
\end{eqnarray}
where $\frac{D}{a}$ is the DM interaction in the continuum limit.

\subsection*{Discussion }

The result by using the $CP^{1}$ model at high temperatures is consistent
with the quantitative analysis of a single triangle in the lattice.
When we expand the average of the TC at high temperature limit, we
find that the $\frac{1}{T^{4}}$ order has the opposite sign of the $\frac{1}{T^{3}}$
 order term and is proportional to $JD^{2}B$. We employed the Pauli-Villars
regularization to calculate the effective action including the quadratic
terms of $\alpha_{i}$, since we need to prove that it is $U(1)$ gauge invariant
and find $\alpha_{i}(q_{i}q_{j}-q^{2}\delta_{ij})\alpha_{j}$ term,
which is the $b^{2}$ term. The momentum cutoff scheme is not proper
for $S_{\alpha^{2}}$, since  it breaks the gauge symmetry, but $S_{hb}$
is gauge invariant because $h$ and $b$ are gauge invariant.
In finite size systems, there exists the bound of momentum. Without
considering the gauge invariance of $hb$ terms, the momentum cutoff scheme
is applied to obtain the effective action of $S_{hb}$. We replace
$\Lambda_{PV}$ by $\Lambda$ for the approximation of the TC because
we only need to know how the TC evolves with the temperature. Also
in momentum cutoff scheme, we do not need to assume $m_{0}^{2}\ll\Lambda^{2}$, so
we can extend the temperature region, since it is not limited by $f\ll1$, and the 
temperature constraint only comes from the large $\mathcal{N}$ approximation
of solving the saddle point equation. This is the reason why we can
perform this model in a relatively high temperature region.

\end{document}